# Hybrid thin-film lithium niobate micro-ring acousto-optic modulator for microwave-to-optical conversion

Lei Wan[1,5,#,*], Jiying Huang[2,#], Meixun Wen[2,#], Huan Li[3,*], Wenfeng Zhou[4], Zhiqiang Yang[5], Yuping Chen[1,4], Huilong Liu[1], Siqing Zeng[5], Dong Liu[5], Shuixian Yang[5], Daoxin Dai[3], Zhaohui Li[5,6,*]

[1]School of Physics, Ningxia University, Yinchuan 750021, China
[2]Department of Electronic Engineering, College of Information Science and Technology, Jinan University, Guangzhou 510632, China
[3]State Key Laboratory of Extreme Photonics and Instrumentation, Center for Optical & Electromagnetic Research, College of Optical Science and Engineering, International Research Center for Advanced Photonics, Zhejiang University, Zijingang Campus, Hangzhou 310058, China
[4]State Key Laboratory of Advanced Optical Communication Systems and Networks, School of Physics and Astronomy, Shanghai Jiao Tong University, Shanghai 200240, China
[5]Guangdong Provincial Key Laboratory of Optoelectronic Information Processing Chips and Systems, Sun Yat-sen University, Guangzhou 510275, China
[6]Southern Marine Science and Engineering Guangdong Laboratory (Zhuhai), Zhuhai 519000, China
[#]These authors contributed equally to this work
*wanlei@nxu.edu.cn
*lihuan20@zju.edu.cn
*lzhh88@sysu.edu.cn

**Abstract**: Highly efficient acousto-optic modulation plays a vital role in the microwave-to-optical conversion. Herein, we demonstrate a hybrid thin-film lithium niobate (TFLN) racetrack micro-ring acousto-optic modulator (AOM) implemented with low-loss chalcogenide (ChG) waveguide. By engineering the electrode configuration of the interdigital transducer, the double-arm micro-ring acousto-optic modulation is experimentally confirmed in nonsuspended ChG loaded TFLN waveguide platform. Varying the position of blue-detuned bias point, the half-wave-voltage-length product $V_\pi L$ of the hybrid TFLN micro-ring AOM is as small as 9 mV·cm. Accordingly, the acousto-optic coupling strength is estimated to be 0.48 Hz s$^{1/2}$ at acoustic frequency of 0.84 GHz. By analyzing the generation of phonon number from the piezoelectric transducer, the microwave-to-optical conversion efficiency is calculated to be 0.05%, approximately one order of magnitude larger than that of the state-of-the-art suspended counterpart. Efficient microwave-to-optical conversion thus provides new opportunities for low-power-consumption quantum information transduction using the TFLN-ChG hybrid piezo-optomechanical devices.

## Introduction

Integrated acousto-optics as an emerging field are exciting great interests in the general community of photonic integrated circuits because it bridges the conversion between the classical and quantum information [1-5]. Highly efficient microwave-to-optical conversion is crucial for the integrated microwave signal processing, quantum qubit transmission and optomechanical sensing [6-12]. An on-chip acousto-optic modulator (AOM) as a key element is indispensable for efficient microwave-to-optical conversion, driving the photon-phonon interaction in monolithic piezo-optomechanical devices. The realization of high-performance integrated AOM is closely dependent on the piezoelectric and photoelastic effects of waveguide materials. Different AOMs have been demonstrated based on typical piezoelectric materials such as aluminum nitride (AlN) [13-16], aluminum scandium nitride (AlScN) [17-21], gallium arsenide (GaAs) [22-24], gallium phosphide (GaP) [25-27], and gallium nitride (GaN) [28-31]. However, the modulation performance of these devices still has large space to improve for satisfying the novel scenarios in microwave or optical signal processing.

With the development of thin-film lithium niobate (TFLN) photonics, the study of high-performance integrated AOMs is pushed to the new level due to its superior piezoelectric and photoelastic properties [32-34]. Generally, the suspended TFLN structure is preferentially considered to realize highly efficient AOM based on homogeneous TFLN platform. Compared with the Mach-Zehnder interferometers (MZIs), racetrack (RT) cavities-based AOMs are demonstrated with higher modulation efficiencies. Specifically, using a 100 μm-long interdigital transducer (IDT) loaded suspended acoustic resonator, a RT AOM is proposed based on TFLN waveguide with a state-of-the-art half-wave-voltage-length product ($V_{\pi}L$) of 7.7 mV·cm [35]. Almost two orders of magnitude improvement are realized compared with the nonsuspended counterpart. The accurate fabrication of suspended acoustic resonator increases the difficulty of device implementation, and reduces the power handling capacity of device. To further improve the modulation efficiency of RT AOM in nonsuspened TFLN platform, we propose a hybrid-integration device prototype based on chalcogenide (GhG) loaded lithium niobate on insulator (LNOI) platform to increase the practicability and robustness of device for microwave-to-optical conversion. Single-arm-modulated RT AOM has been verified with a $V_{\pi}L$ of 20 mV·cm in our previous work [36]. Double-arm TFLN-ChG hybrid RT AOM is yet to be demonstrated. Theoretical analysis about optomechanical coupling characteristics is unclear in current references with respect to the RT AOM without acoustic resonator. Accordingly, the microwave-to-optical conversion needs to be clarified using aforementioned RT AOM configuration.

In this work, we demonstrate a highly efficient RT AOM based on nonsuspended ChG loaded LNOI platform. The principle of RT AOM is systemically introduced to reveal the photon-phonon interaction. To illustrate the evolution of RT AOM, single-arm and double-arm configurations are presented together through the engineering of IDT. The displacement distributions of excited surface acoustic waves (SAWs) in two straight waveguides of the RT resonator are carefully designed to compare the modulation characteristics between the symmetric and antisymmetric acoustic modes. Variations of the modulation efficiencies of a double-arm RT AOM at the different bias points are investigated for three representative acoustic modes. As a proof, the microwave-to-optical conversion are demonstrated using efficient double-arm RT AOMs, showing their tremendous potentials in high-performance microwave signal processing.

## Results
### Device Design
Figure 1(a) shows the schematic diagram of the double-arm RT AOM based on nonsuspended ChG loaded LNOI platform. The device consists of the TFLN-ChG hybrid micro-ring resonator and the incorporated IDT with even number of electrodes. The hybrid micro-ring resonator is employed to confine optical resonance mode, forming a high Q-factor microcavity. The IDT is used to generate symmetric SAW perturbing optical resonance mode by photoelastic and moving boundary effects. The perturbation strength is determined by the intrinsic material property and geometry of planar waveguide in principle. To increase the phase modulation of the confined optical mode in the hybrid waveguide, the developed ChG $Ge_{25}Sb_{10}S_{65}$ membrane with large photoelastic coefficient is integrated over the X-cut TFLN slab with thickness of 400 nm, creating the core layer of the hybrid waveguide together.

In general, the slope detection mechanism is primarily chosen to transduce the perturbations from the SAWs. Figure 1(b) presents the schematic diagram of the optical transmission curve of RT AOM. Once the SAWs pass through the straight waveguides in the hybrid RT resonator, the variation of the optical power at a specific bias point is proportional to the slope of the transmission curve, as illustrated in Fig. 1(c). Corresponding optical power changes are recorded by a photo receiver and subsequently converted into opto-acoustic $S_{21}$ parameter using a vector network analyzer (VNA) (as denoted by the orange line). To compare the modulation performance with MZI AOM, we theoretically define an effective $V_\pi$ to reflect phase modulation. The relative formula is described as:

$$V_\pi = \frac{\pi R_{PD} P_{opt\_0}}{S_{21}} \cdot \frac{\left.\dfrac{dT}{d\phi}\right|_{\phi=\phi_L}}{T(\phi_L)} \ , \tag{1}$$

where $R_{PD}$ is the responsivity of a photo receiver, and $P_{opt\_0}$ is the DC optical output power. $T$ is the normalized transmission of a micro-ring resonator, and $dT/d\phi$ is the slope of a normalized transmission curve. The detailed derivation is illustrated in Supplementary note 1. An appropriate bias point is thus critical to achieve highly efficient acousto-optic modulation exploiting a micro-ring configuration.

To confirm the conception of the device, the fundamental transverse electric (TE) mode is designed for the acousto-optic modulation, as shown in Fig. 2(a). Herein, the width of ChG rectangular waveguide is denoted as $W$, and the height is denoted as $H$. Most of the optical energy is confined into the ChG waveguide so as to fully utilize its excellent photoelastic property. The dependence of the acousto-optic overlap factors on the variations of the geometries of the ChG waveguide is theoretically calculated to estimate photon-phonon interaction in hybrid RT AOM (see Supplementary note 2), as shown in Fig. 2(b). When $W$ and $H$ are around 1.4 μm and 1.1 μm, respectively, the acousto-optic overlap factor of up to 0.5 can be ideally obtained. The optimal dimensions are attributed to fundamental TE mode distribution approaching to the strain field of half acoustic wavelength. However, considering the practical propagation loss, 2 μm-width and 0.85 μm-height ChG waveguide is chosen for hybrid waveguide. The corresponding acousto-optic overlap factor is 0.18, and 83% optical energy is confined in the ChG waveguide, as shown in Fig. 2(a). In the simulation, the IDT with 50 pairs of electrodes is embedded into the center between both straight waveguides to

excite the symmetric acoustic mode. The wavelength of the acoustic wave is set to be

3.2 μm, corresponding to an electrode width of 800 nm. The normalized $S_{xx}$ and $S_{yy}$ strain components at the frequency of 0.8413 GHz show the same displacement distributions in both straight waveguides along the crystal Z direction, predicting the double-arm phase modulation in modulator, as presented in Fig. 2(c) & 2(d). Given that $S_{yy} \approx 2S_{xx}$, out-of-plane strain dominates the deformation of the hybrid waveguide. Similar displacement distributions in both straight waveguides are also demonstrated at higher acoustic frequencies for the same IDT configuration, which is explained in the Supplementary note 3.

**Device characterization**

Figure 3(a) shows the optical microscope image of a fabricated double-arm RT AOM, and zoomed-in pulley coupling region is presented in Fig. 3(c). To realize good contact with G-S-G RF probe, two-section IDT contacted by the three pads is designed and fabricated. Figure 3(b) shows the scanning electron microscope (SEM) image of partial IDT, and corresponding zoomed-in image is presented in Fig. 3(d). Figure 4(a) shows the measurement system of a RT AOM, as illustrated in our previous work [37].

**Acousto-optic modulation comparison**

To reasonably compare the modulation efficiency between single-arm and double-arm configurations, we integrate both IDT configurations with a single RT micro-ring resonator to share the same transmission curve, as shown in Fig. 4(b). The loaded Q-factor is calculated to be $2.1 \times 10^5$ through the Lorentzian fitting. Figure 4(c) shows the measurement results for the reflection coefficient spectra $S_{11}$ of the IDTs and the transmission coefficient spectra $S_{21}$ of the modulators. Due to the same structure parameters of IDTs, similar valleys profile is experimentally obtained in the $S_{11}$, indicating three acoustic modes with frequencies of 0.85 GHz, 1.48 GHz, and 1.52 GHz, respectively, denoted as 1, 2, and 3. Compared with the acoustic mode 1, acoustic modes 2 and 3 with $1-|S_{11}|^2$ of 99.9% and 95.6% have better microwave transmissions. However, the value of $S_{21}$ (-45.67 dB) of the double-arm RT AOM corresponding to the acoustic mode 1 is significantly better than that corresponding to the acoustic modes 2 and 3 (-57.01 dB, -57.27 dB), due to its larger acousto-optic overlap factor. For the single-arm RT AOM, no obvious peak appears in the $S_{21}$ corresponding to the mode 3, indicating that the fabricated double-arm RT AOM has superior modulation efficiency in comparison with single-arm counterpart. Specifically, the $V_\pi L$ of the double-arm RT AOM is estimated to be 15 mV·cm at 0.85 GHz, approximately two times higher than that of the single-arm counterpart, demonstrating the double-arm modulation effect due to the symmetric SAW perturbations as well.

The efficiency of the RT AOM is also determined by the IDT design. To reveal the effects of the number of electrodes on the modulation efficiency, we compare the modulation performance of two RT AOMs with 120.5 and 120 pairs of embedded electrodes, respectively. Figures 5(a) & 5(c) show the representative transmission spectra of two double-arm RT AOMs with comparable Q-factors and extinction ratios. For Device B, the values of $S_{21}$ at different acoustic modes (as shown in Fig. 5(b)) are significantly lower than that of Device C with 120 pairs of electrodes under the similar DC bias points (-26 dBm) (as presented in Fig. 5(d)). For the dominated acoustic mode 1, the modulation efficiency $V_\pi L$ of Device B is measured to be 26.4 mV·cm, nearly three times higher than value of Device C. Accordingly, the values of $V_\pi L$ of Device B at the acoustic mode 2 and 3 are estimated to be 166 mV·cm and 181 mV·cm, respectively, almost 1.5 times higher than those of Device C at the same frequencies. It is attributed to that the opposite acoustic strain distributions ($S_{xx}$ and $S_{yy}$) are generated within the two straight waveguides of Device B, leading to partial phase change cancelled. The detailed acoustic strain components are provided in Supplementary note 4. Additionally, the theoretical effective refractive index changes of both devices for different acoustic modes are also simulated to illustrate the difference of phase modulation under the two IDT configurations (see Supplementary note 2). Practical deviations in modulation efficiency may be caused by the scattering of the ChG waveguides for SAWs.

Based on the above proposed double-arm RT AOM with an IDT of 120 pairs of electrodes, we further investigate the effect of the bias points in the transmission curve on the modulation efficiency under the three representative acoustic modes, as shown in Fig. 6(a). Figure 6(b) shows the variations of modulation efficiencies of Device C at different blue-detuned bias points for the three acoustic modes. It is evident that the modulation efficiency of Device C is decreased with the increase of acoustic frequency. As explained before, the acousto-optic overlap factor of acoustic mode at high frequency is dramatically reduced, due to more acoustic energy leaking into the substrate without photon-phonon interaction. However, with the blue-detuned bias point approaching to the center resonance wavelength, the measured modulation efficiencies of device are gradually improved to maximum, consisting well with the theoretical calculation results from Lorentzian fitting (see Supplementary note 1). The achievement of maximized modulation efficiency at a specific acoustic mode corresponds to the evolution of the slope in the transmission curve, verifying the theoretical prediction in working principle. To more clearly present the modulation performance of Device C, we characterize the spectrum of modulated optical sidebands near the minimized $V_\pi L$ of 9 mV·cm at RF power of 12 dBm, as shown in Fig. 6(c). Up to five-order optical sideband is experimentally demonstrated while the theoretical values are accordingly calculated to compare with the experimental result (see Supplementary note 5). Similar evolution trends between both data fully reveal the good modulation capability of proposed double-arm RT AOM.

**Description of microwave-to-optical conversion**
To estimate the performance of proposed RT AOM in microwave-to-optical conversion, the photon number conversion efficiencies $\eta_{\_blue}$ from the microwave frequency to optical sideband frequency as another key figure-of-merit (FOM) is measured at different RF powers for both bias points, as shown in Fig. 6(d). Slight variations of the microwave-to-optical conversion efficiencies at a specific bias point is caused by the changes of transmission curves of the RT resonator with the increase of RF power [38]. In particular, the maximized microwave-to-optical conversion efficiency is calculated to be 0.05% at 0.84 GHz for the bias point of -23 dBm, which is approximately one order of magnitude larger than that of counterpart given in Ref. 41. It is because that the generated phonon number in our device is more than that in suspended modulator. The detailed calculation processes are shown in Supplementary note 6.

## Discussion

Table 1 compares the performances of our proposed hybrid-integration double-arm RT AOM with the state-of-the-art TFLN micro-ring AOMs. Different photonic material platforms have been investigated to engineer the SAW-assisted monolithic micro-ring AOM such as AlN, Si and LN. Thanks to the excellent piezoelectric effect and suspended acoustic resonator, the homogeneous TFLN-based RT AOM has obtained maximized modulation efficiency of 7.7 mV·cm and phase shift per unit square root of microwave power $\alpha_p$ of 1.83 rad mW$^{-1/2}$. Compared with the merits of this device, our developed hybrid RT AOM based on the ChG loaded LNOI platform shows comparable FOMs. The $V_\pi L$ of 9 mV·cm and $\alpha_p$ of 1.49 rad mW$^{-1/2}$ are firstly reported to research community employing nonsuspended waveguide platform without acoustic resonator, seriously simplifying the fabrication recipe of high-performance AOM. In addition, the defined acousto-optic coupling coefficient $g'/2\pi$ of our device is calculated to be 0.48 Hz s$^{1/2}$ between the acoustic Rayleigh mode at 0.84 GHz and fundamental TE optical mode (see Supplementary note 7).

Demonstration of the highly efficient RT AOM based on the TFLN-ChG hybrid waveguide provides new pathway to study novel on-chip piezo-optomechanical interaction devices. In addition, high-performance microwave photon filter and optical isolator are also expected to present superior results based on the developed device prototype. In the future, the introduction of guided acoustic modes in the hybrid waveguide platform will provide much more opportunities to dramatically improve performance of the hybrid-integration AOMs, further expanding innovative applications in optical analog computing[42], LiDAR[43], quantum information transduction[44].

In conclusion, we propose a hybrid-integration RT AOM on ChG loaded LNOI platform. Combing with the slope detection mechanism of a micro-ring resonator, effective $V_\pi$ of the RT AOM is theoretically given to estimate the performance of device. By engineering the electrode number of the IDT inserted within the RT resonator, the symmetric Rayleigh acoustic mode appearing in the two straight waveguides is excited to form double-arm modulation configuration. Compared with the single-arm modulator, the double-arm AOM obtains approximately two times improvement in modulation efficiency. As a result, a hybrid RT AOM with modulation efficiency up to 9 mV·cm is experimentally confirmed by optimizing the optical waveguide and IDT. The effect of DC bias point in the transmission curve on the modulation efficiency is analyzed to reveal the evolution of modulation efficiency. To reflect the capability of the highly efficient RT AOM in microwave-to-optical conversion, the acousto-optic coupling coefficient and photon number conversion efficiency of the device is calculated to be 0.48 Hz s$^{1/2}$ and 0.05%, respectively, at 0.84 GHz.

## Materials and Methods
### Device fabrication
The devices were fabricated on an X-cut thin-film LNOI wafer purchased from NANOLN, where the nominal thickness of the LN layer was 400 nm. We first deposited an 850 nm-thick $Ge_{25}Sb_{10}S_{65}$ membrane on the LNOI wafer by the thermal evaporation method. Then, we performed electron-beam lithography (EBL) to design the RT micro-ring structure as a mask using an electron-beam resist (ARP 6200.13) and transferred the photonic waveguide onto the $Ge_{25}Sb_{10}S_{65}$ film using reactive ion etching. Finally, the IDTs were fabricated through a lift-off process involving second-step EBL and gold deposition, where the thickness of the gold electrodes was 100 nm, with a 10 nm Ti adhesive layer that was previously deposited. The thickness of ARP6200.13 was controlled to 400 nm during spin coating using a spinning speed of 4000 rpm.

### Measurement methods
A continue wave (CW) tunable laser (Keysight 81940A) was used to measure the transmission of micro-ring resonator. The characterization of the $S_{11}$ spectra for the IDTs was conducted using a VNA (Keysight, N5225A) with the aid of a microwave probe (GGB, 40A-GSG-100-DP). By choosing a proper bias wavelength in the transmission spectrum of the RT micro-ring resonator, the $S_{21}$ spectrum could be obtained by scanning the microwave frequency in the VNA when the modulated optical wave was converted into an electrical signal via a high-speed photodiode (Newport, 1544-B). The spectrum of the modulation sidebands was recorded by connecting the output fiber to a high-precision OSA (APEX, AP2088A).

### Acknowledgments
We acknowledge the funding support provided by the National Natural Science Foundation of China (Grant Nos. 62175095, 62335014, 12134009), the Open Project of State Key Laboratory of Advanced Optical Communication Systems and Networks in Shanghai Jiao Tong University (No. 2024GZKF002).

### Author contributions
L.W., Z.L., J.H., and M.W. conceived the device design. S.Z., D.L., and S.Y. carried out the device fabrication. M.W., W.Z., and Z.Y. performed the device measurements. L.W., J.H., and H.L. carried out the data analysis. All authors contributed to the writing. L.W. finalized the paper. L.W. and Z.L. supervised the project.

### Conflict of Interest
The authors declare no conflicts of interests.

### References
1. D. Marpaung, J. Yao, and J. Capmany, "Integrated microwave photonics," Nat. Photonics **13**, 80-90 (2019).
2. A. Rueda, F. Sedlmeir, M. C. Collodo et al., "Efficient microwave to optical photon conversion: an electro-optical realization," Optica **3**, 597-604 (2016).

3.    K. Stannigel, P. Rabl, A. S. Sørensen et al., "Optomechanical transducers for long-distance quantum communication," Phys. Rev. Lett. **105**, 220501 (2010).

4.    M. Tsang, "Cavity quantum electro-optics," Phys. Rev. A **81**, 063837 (2010).

5.    G. Wendin, "Quantum information processing with superconducting circuits: a review," Reports on Progress in Physics **80**, 106001 (2017).

6.    I. S. Amiri, and A. Afroozeh *Ring resonator systems to perform optical communication enhancement using soliton* (Springer, 2014).

7.    J. Capmany, and D. Novak, "Microwave photonics combines two worlds," Nat. Photonics **1**, 319 (2007).

8.    K. Fang, M. H. Matheny, X. Luan et al., "Optical transduction and routing of microwave phonons in cavity-optomechanical circuits," Nat. Photonics **10**, 489-496 (2016).

9.    E. Gavartin, P. Verlot, and T. J. Kippenberg, "A hybrid on-chip optomechanical transducer for ultrasensitive force measurements," Nat. Nanotechnol. **7**, 509-514 (2012).

10.    R. Riedinger, A. Wallucks, I. Marinković et al., "Remote quantum entanglement between two micromechanical oscillators," Nature **556**, 473-477 (2018).

11.    L. Zhang, C. Cui, P.-K. Chen et al., "Integrated-waveguide-based acousto-optic modulation with complete optical conversion," Optica **11** (2024).

12.    Z. Tao, B. Shen, W. Li et al., "Versatile photonic molecule switch in multimode microresonators," Light: Science & Applications **13** (2024).

13.    A. Cleland, M. Pophristic, and I. Ferguson, "Single-crystal aluminum nitride nanomechanical resonators," Appl. Phys. Lett. **79**, 2070-2072 (2001).

14.    L. Fan, X. Sun, C. Xiong et al., "Aluminum nitride piezo-acousto-photonic crystal nanocavity with high quality factors," Appl. Phys. Lett. **102**, 153507 (2013).

15.    H. Li, S. A. Tadesse, Q. Liu et al., "Nanophotonic cavity optomechanics with propagating acoustic waves at frequencies up to 12 GHz," Optica **2**, 826-831 (2015).

16.    C. Xiong, W. H. Pernice, X. Sun et al., "Aluminum nitride as a new material for chip-scale optomechanics and nonlinear optics," New Journal of Physics **14**, 095014 (2012).

17.    M. Akiyama, K. Umeda, A. Honda et al., "Influence of scandium concentration on power generation figure of merit of scandium aluminum nitride thin films," Appl. Phys. Lett. **102**, 021915 (2013).

18.    A. Ding, L. Kirste, Y. Lu et al., "Enhanced electromechanical coupling in SAW resonators based on sputtered non-polar Al0. 77Sc0. 23N 11 2⁻ 0 thin films," Appl. Phys. Lett. **116**, 101903 (2020).

19.    R. Matloub, M. Hadad, A. Mazzalai et al., "Piezoelectric Al1− xScxN thin films: A semiconductor compatible solution for mechanical energy harvesting and sensors," Appl. Phys. Lett. **102**, 152903 (2013).

20.    K. Men, H. Liu, X. Wang et al., "AlScN films prepared by alloy targets and SAW device characteristics," Journal of Rare Earths **41**, 434-439 (2023).

21.    W. Wang, P. M. Mayrhofer, X. He et al., "High performance AlScN thin film based surface acoustic wave devices with large electromechanical coupling coefficient," Appl. Phys. Lett. **105**, 133502 (2014).

22.    K. C. Balram, M. Davanço, J. Y. Lim et al., "Moving boundary and photoelastic coupling in GaAs optomechanical resonators," Optica **1**, 414-420 (2014).

23.    S. Combrié, A. De Rossi, Q. V. Tran et al., "GaAs photonic crystal cavity with ultrahigh

Q: microwatt nonlinearity at 1.55 μm," Opt. Lett. **33**, 1908-1910 (2008).

24. G. Shambat, B. Ellis, M. A. Mayer et al., "Ultra-low power fiber-coupled gallium arsenide photonic crystal cavity electro-optic modulator," Opt. Express **19**, 7530-7536 (2011).

25. S. Hönl, Y. Popoff, D. Caimi et al., "Microwave-to-optical conversion with a gallium phosphide photonic crystal cavity," Nat. Commun. **13**, 2065 (2022).

26. M. Mitchell, A. C. Hryciw, and P. E. Barclay, "Cavity optomechanics in gallium phosphide microdisks," Appl. Phys. Lett. **104**, 141104 (2014).

27. K. Schneider, Y. Baumgartner, S. Hönl et al., "Optomechanics with one-dimensional gallium phosphide photonic crystal cavities," Optica **6**, 577-584 (2019).

28. H. D. Jabbar, M. A. Fakhri, and M. J. AbdulRazzaq, "Gallium nitride–based photodiode: a review," Materials Today: Proceedings **42**, 2829-2834 (2021).

29. C. Lueng, H. L. Chan, C. Surya et al., "Piezoelectric coefficient of aluminum nitride and gallium nitride," J. Appl. Phys. **88**, 5360-5363 (2000).

30. M. Rais-Zadeh, V. J. Gokhale, A. Ansari et al., "Gallium nitride as an electromechanical material," Journal of Microelectromechanical Systems **23**, 1252-1271 (2014).

31. Y. Zheng, C. Sun, B. Xiong et al., "Integrated gallium nitride nonlinear photonics," Laser & Photonics Reviews **16**, 2100071 (2022).

32. B. Desiatov, A. Shams-Ansari, M. Zhang et al., "Ultra-low-loss integrated visible photonics using thin-film lithium niobate," Optica **6**, 380-384 (2019).

33. W. Jiang, C. J. Sarabalis, Y. D. Dahmani et al., "Efficient bidirectional piezo-optomechanical transduction between microwave and optical frequency," Nat. Commun. **11**, 1166 (2020).

34. A. J. Mercante, S. Shi, P. Yao et al., "Thin film lithium niobate electro-optic modulator with terahertz operating bandwidth," Opt. Express **26**, 14810-14816 (2018).

35. L. Shao, M. Yu, S. Maity et al., "Microwave-to-optical conversion using lithium niobate thin-film acoustic resonators," Optica **6**, 1498-1505 (2019).

36. Z. Yang, M. Wen, L. Wan et al., "Efficient acousto-optic modulation using a microring resonator on a thin-film lithium niobate–chalcogenide hybrid platform," Opt. Lett. **47**, 3808-3811 (2022).

37. L. Wan, Z. Yang, W. Zhou et al., "Highly efficient acousto-optic modulation using nonsuspended thin-film lithium niobate-chalcogenide hybrid waveguides," Light: Science & Applications **11**, 145 (2022).

38. L. Cai, A. Mahmoud, M. Khan et al., "Acousto-optical modulation of thin film lithium niobate waveguide devices," Photonics Research **7** (2019).

39. S. A. Tadesse, and M. Li, "Sub-optical wavelength acoustic wave modulation of integrated photonic resonators at microwave frequencies," Nature Communications **5** (2014).

40. D. Munk, M. Katzman, M. Hen et al., "Surface acoustic wave photonic devices in silicon on insulator," Nature Communications **10** (2019).

41. L. Shao, M. Yu, S. Maity et al., "Microwave-to-optical conversion using lithium niobate thin-film acoustic resonators," Optica **6** (2019).

42. M. Dong, G. Clark, A. J. Leenheer et al., "High-speed programmable photonic circuits in a cryogenically compatible, visible–near-infrared 200 mm CMOS architecture," Nature Photonics **16**, 59-65 (2021).

43. B. Li, Q. Lin, and M. Li, "Frequency–angular resolving LiDAR using chip-scale acousto-

optic beam steering," Nature **620**, 316-322 (2023).

44. Y. Wang, J. Lee, and P. X. L. Feng, "Perspectives on phononic waveguides for on-chip classical and quantum transduction," Applied Physics Letters **124** (2024).

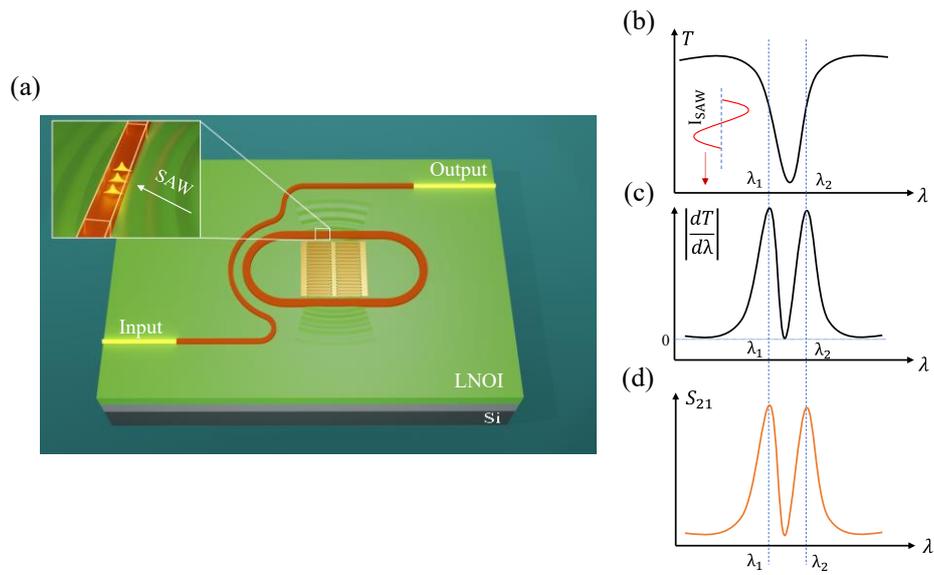

**Fig. 1** Schematic diagram and working principle of the double-arm RT AOM. (a) Device prototype of a RT AOM. (b) Typical optical transmission curve of a device, and (c) corresponding slope variation curve. (d) Evolution of $S_{21}$ under the perturbation of acoustic wave.

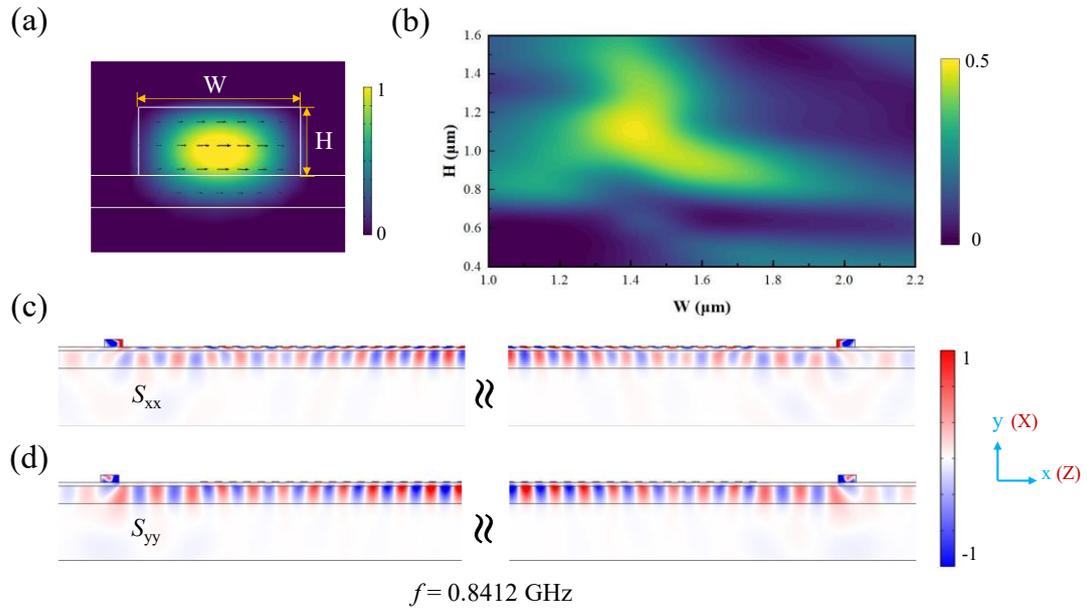

**Fig. 2** Optical and acoustic modes simulation results of the double-arm RT AOM. (a) Normalized fundamental TE mode in the defined hybrid waveguide. (b) Dependence of the acousto-optic overlap factors on the variations of the geometries of hybrid waveguides. Normalized (c) $S_{xx}$ and (d) $S_{yy}$ strain distributions in both straight waveguides of double-arm RT AOM at the frequency of 0.8412 GHz.

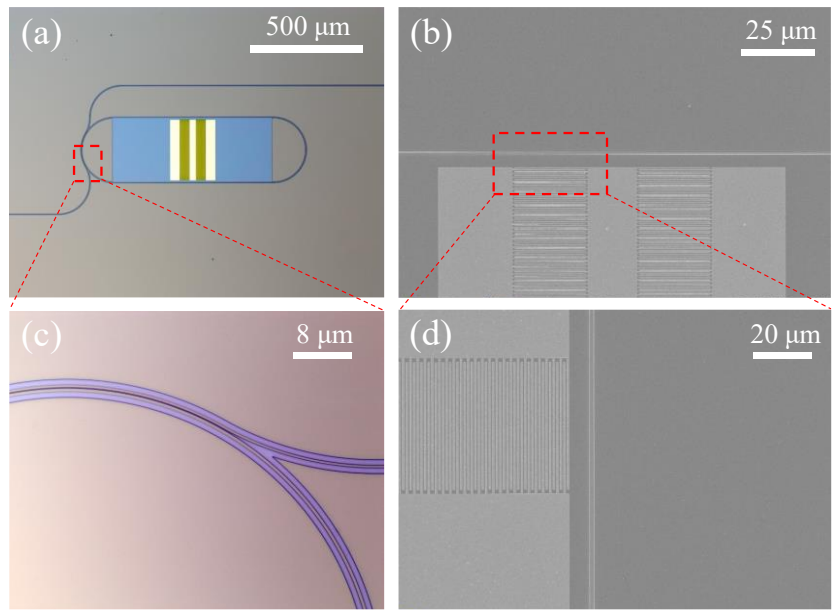

**Fig. 3** Images of a fabricated double-arm RT AOM. (a) Optical microscope image of a device and (c) corresponding zoomed-in image for partial coupling region. (b) SEM image of an embedded IDT and (d) corresponding zoomed-in image.

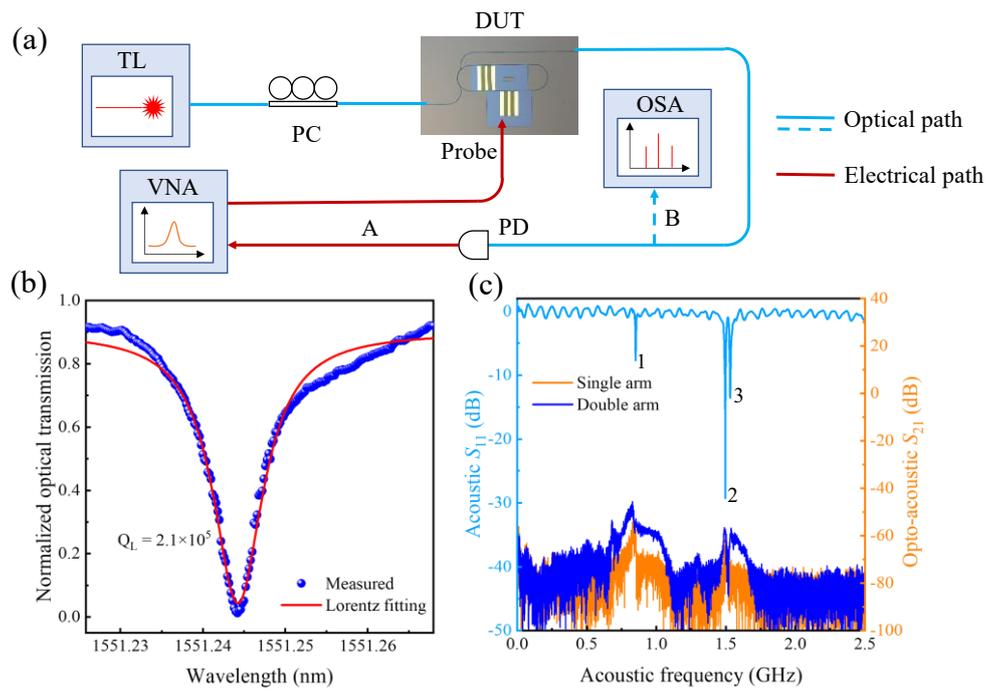

**Fig. 4** Characterization results of an integrated RT AOM (Device A). (a) Measured transmission curve of the device. (b) Measured $S_{11}$ and $S_{21}$ spectra corresponding to single-arm and double-arm configurations.

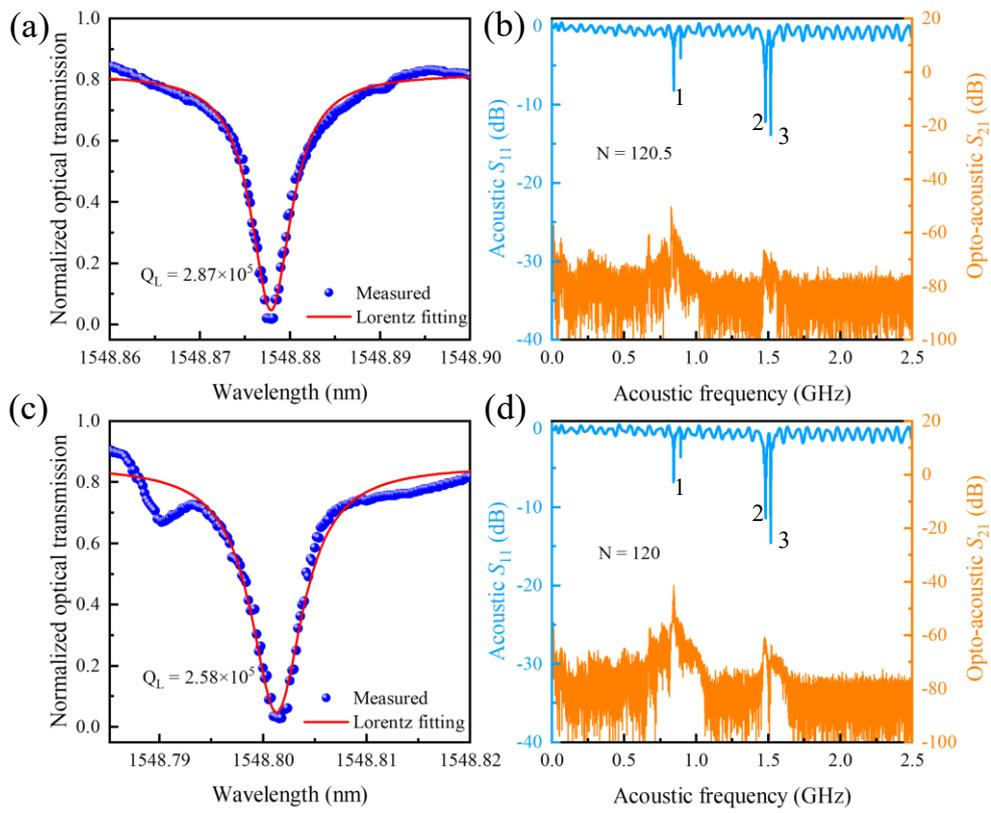

**Fig. 5** Characteristics comparison of two double-arm RT AOMs with different embedded IDT configurations (Device B and Device C). (a) Measured transmission curve of Device B and (b) corresponding $S_{11}$ ($S_{21}$) spectra. (c) Measured transmission curve of Device C and (d) corresponding $S_{11}$ ($S_{21}$) spectra.

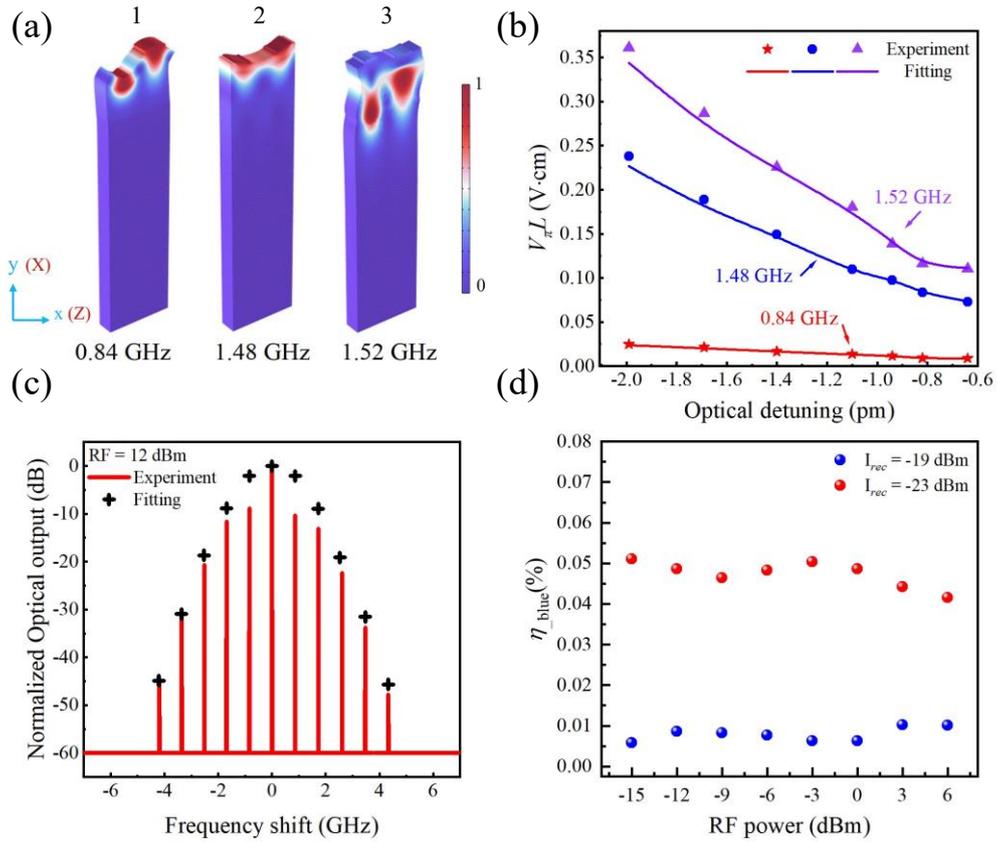

**Fig. 6** Effects of the DC bias points on the modulation efficiencies of the double-arm RT AOM (Device C). (a) Normalized displacement profiles for the three representative acoustic modes. (b) Variations of the modulation efficiencies of Device C at different bias points. (c) Measured spectrum and theoretical fitting result of modulated optical sidebands at RF power of 12 dBm near the minimized $V_\pi L$ at 0.84 GHz. (d) Variations of microwave-to-optical conversion efficiencies at different RF powers for both bias points.

**Table 1. Comparison of modulation metrics for TFLN RT AOMs**

| Ref. | Platform | Acoustic cavity | Frequency (GHz) | $1-|S_{11}|^2$ (%) | $L$ (µm) | $Q_L$ | $\alpha_p$ (rad mW$^{-1/2}$) | $V_\pi L$ (V·cm) | $g'/2\pi$ (Hz s$^{1/2}$) | $\eta$ (%) |
|---|---|---|---|---|---|---|---|---|---|---|
| [39] | AlN | x | 10.6 | 60.2 | / | $8\times10^4$ | / | / | / | / |
| [38] | LN | √ | 0.11 | 42 | 2400 | $2.66\times10^5$ | / | / | / | / |
| [40] | SOI | x | 2.4 | / | 60 | $4\times10^4$ | / | / | / | / |
| [41] | LN | √ | 2.0 | 64 | 100 | $2.2\times10^6$ | 1.83 | 0.0077 | 1.97 | 0.0017 |
| [36] | ChG/LN | x | 0.84 | 98.4 | 120 | $5\times10^5$ | 0.57 | 0.02 | 0.037 | / |
| **This work** | GhG/LN | x | 0.84 | 79.1 | 120 | $3.71\times10^5$ | 1.49 | 0.009 | 0.48 | 0.05 |
| | GhG/LN | x | 1.48 | 92.6 | 120 | $3.71\times10^5$ | 0.15 | 0.083 | 0.063 | / |
| | GhG/LN | x | 1.52 | 96.5 | 120 | $3.71\times10^5$ | 0.11 | 0.12 | 0.047 | / |